\documentclass[a4paper,12pt]{article}
\usepackage[top=3cm,bottom=3.4cm,left=2cm,right=2cm]{geometry}

\usepackage{amsmath,amssymb,graphicx}

\makeatletter
\def\slash#1{{\ooalign{\hfil/\hfil\crcr$#1$}}}

\def\braket<#1|#2|#3>{%
\gdef\@bra{#1}\gdef\@opr{#2}\gdef\@ket{#3}\@braket}
\def\bracket#1{\expandafter\bunri#1\@nil\relax\@braket}
\def\bunri#1|#2|#3\@nil{%
\gdef\@bra{#1}\gdef\@opr{#2}\gdef\@ket{#3}}
\def\@braket{%
\setbox0=\hbox{$\displaystyle {\@bra}{\@ket}$}
\dimen0\ht0 \dimen1\dp0
\setbox0=\hbox{$\displaystyle {\@opr}$}
\@tempdima=.8\ht0 \@tempdimb=.8\dp0
\ifdim\@tempdima>\dimen0\relax\dimen0=\@tempdima\fi
\ifdim\@tempdimb>\dimen1\relax\dimen1=\@tempdimb\fi
\setbox1=\hbox{\vrule height\dimen0 depth\dimen1 width\z@}
\def\Strut{\relax\ifmmode\copy1\else\unhcopy1\fi}
\def\sp@ce{\kern.09em}
\ifx\@bra\empty\relax
 \ifx\@opr\empty\relax \left|\Strut\sp@ce{\@ket}\right>
 \else 
   \ifx\@ket\empty\relax \left<{\@opr}\right> 
   \else {\@opr}\left|\Strut\sp@ce{\@ket}\right> \fi
 \fi
\else
 \ifx\@opr\empty\relax
   \ifx\@ket\empty\relax \left<\Strut{\@bra}\sp@ce\right|
   \else \left<\Strut{\@bra}\sp@ce\right|\left.
                         \kern-.22em\Strut{\@ket}\right> \fi
 \else 
   \ifx\@ket\empty\relax%
         \left<\Strut{\@bra}\sp@ce\right|{\@opr}
   \else \left<\Strut{\@bra}\sp@ce\right|{\@opr}
                \left|\Strut\sp@ce{\@ket}\right> \fi
 \fi
\fi}
\newbox\scb@x \newbox\scscb@x
\newdimen\shaderule \newdimen\@scw
\newdimen\@savetbaselineshift
\newdimen\@saveybaselineshift
\def\@hobox#1#2#3{\hbox to\@scw{\@oval(#3,#3)[#1]\hfil\@oval(#3,#3)[#2]}}
\def\@vrf@#1#2{\vskip#1\leaders\vrule width#2\vfil\vskip#1}
\def\screen{\@ifnextchar[{\@screen}{\@screen[\linewidth]}}
\def\@screen[#1]{\newdimen\t@mpdim \t@mpdim#1%
\def\@r{20}\def\@s{9.8pt}\def\@sx{10pt}
\ifdim\t@mpdim=\linewidth \par\vspace{.3\baselineskip}\fi
\@scw\t@mpdim \advance\@scw -\@r pt
\setbox\scb@x=\hbox\bgroup\begin{minipage}
{\@scw\tbaselineshift\@savetbaselineshift}
}%
\def\endscreen{\end{minipage}\egroup%
\setbox\scscb@x=\hbox to\t@mpdim{\hfil\vbox{\offinterlineskip%
\thicklines\setlength{\unitlength}{1pt}\hrule height.8pt\kern\@s%
\@hobox{tl}{tr}{\@r}\kern-\@sx\box\scb@x\@hobox{bl}{br}{\@r}%
\kern-.8pt\hrule height.8pt}\hfil}\dimen0=\ht\scscb@x%
\noindent\hbox to\t@mpdim{\hbox to.6pt{\vbox to\dimen0{\@vrf@{\@sx}{.8pt}}%
\hss}\box\scscb@x\hbox to.6pt{\hss\vbox to\dimen0{\@vrf@{\@sx}{.9pt}}}\hss}}%
\def\shadedbox{\@ifnextchar[{\@shadedbox}{\@shadedbox[\linewidth]}}
\def\@shadedbox[#1]{\dimen0#1 \advance\dimen0-20pt 
\advance\dimen0-2\fboxrule \advance\dimen0-\shaderule
\setbox\@tempboxa=\hbox\bgroup\minipage{\dimen0}\vspace{1ex}}
\def\endshadedbox{\vspace{1ex}\endminipage\egroup%
\dimen0=10pt \advance\dimen0-\fboxrule
\setbox\@tempboxa=\hbox{\kern\dimen0\unhbox\@tempboxa\kern\dimen0}%
\setbox0=\vbox{\hrule \@height \fboxrule
\hbox{\vrule \@width \fboxrule \hskip-\fboxrule
\vbox{\vskip\fboxsep \box\@tempboxa\vskip\fboxsep}%
\hskip-\fboxrule\vrule \@width \fboxrule}\hrule \@height \fboxrule}%
\dimen0=\ht0 \advance\dimen0-\shaderule
\hbox{\hbox to \shaderule{\copy0\hss}\kern 0pt
\vrule width\wd0 height0pt depth\shaderule\hskip-\shaderule
\vrule width\shaderule height\dimen0}}
\newdimen\@itemh\newtoks\@iboxstr\newtoks\@iboxpos
\def\@hleader{\leaders\hrule height .8pt\hfil}
\def\itembox{\@ifnextchar[{\@itembox}{\@itembox[c]}}
\def\@itembox[#1]#2{\par\vspace{.3\baselineskip}{\setbox0=\hbox{#2}%
\global\@itemh\ht0\global\advance\@itemh\dp0}%
\vspace{.5\@itemh}\bgroup\@scw\linewidth
\advance\@scw -20pt\@iboxpos={#1}\@iboxstr={#2}%
\setbox\scb@x=\hbox\bgroup\begin{minipage}{\@scw}\vspace*{.5\@itemh}}%
\def\enditembox{\end{minipage}\egroup%
\setbox\scscb@x=\hbox to\linewidth{\hfil\vbox{\offinterlineskip%
\thicklines\setlength{\unitlength}{1pt}%
\hbox to\@scw{\if l\the\@iboxpos\else\@hleader\fi
\ \vbox to.8pt{\vss\hbox{\the\@iboxstr}\vss}\ \if r\the\@iboxpos\else
\@hleader\fi}\kern9.6pt
\@hobox{tl}{tr}{20}\kern-10pt\box\scb@x\@hobox{bl}{br}{20}%
\kern-.8pt\hrule height.8pt}\hfil}\dimen0=\ht\scscb@x%
\noindent\hbox to\linewidth{\hbox to.6pt{\vbox to\dimen0{\@vrf@{10pt}{.8pt}}%
\hss}\box\scscb@x\hbox to.6pt{\hss\vbox to\dimen0{\@vrf@{10pt}{.9pt}}}%
\hss}\egroup}
\def\sheqnarray{\@ifnextchar[{\@sheqnarray}{\@sheqnarray[.8\linewidth]}}
\def\@sheqnarray[#1]{$$\begin{shadedbox}[#1]\vspace{-2.5ex}\begin{eqnarray}}
\def\endsheqnarray{\end{eqnarray}\vspace{-3.5ex}\end{shadedbox}$$}
\def\shbox#1{\newdimen\tempdima
 \setbox0=\hbox{\rule{2eM}{0pt}#1} \tempdima\wd0 
\begin{shadedbox}[\tempdima]#1\end{shadedbox}}

\def\ket#1{\left|{#1}\right\rangle}
\def\VEV#1{\left\langle{#1}\right\rangle}

\def\nn{\nonumber\\}
\newcommand{\abs}[1]{\left|{#1}\right|}
\def\norm#1{\left|\kern-.09em\left|#1\right|\kern-.1em\right|}
\def\calD{{\cal D}}
\def\half{\hbox{\Large ${1\over2}$}}
\def\quarter{\hbox{\Large ${1\over4}$}}
\def\threehalf{\hbox{\Large ${3\over2}$}}
\makeatother

\renewcommand{\title}[1]{%
	\begin{center} \Large \bf #1 \end{center}%
	}
\renewcommand{\author}[2]{%
	\begin{center} { #1}  \vspace{2mm}\\ %
	  {\it #2}%
	\end{center}%
	\addvspace{\baselineskip}%
	}

\begin{document}
\newpage
\setcounter{section}{0}
\setcounter{equation}{0}
\setcounter{figure}{0}
\baselineskip 5mm
\begin{flushright}
MISC-2017-01 \\
February, 2017
\end{flushright}
\vspace{10mm}
\title{%
Spontaneous Supersymmetry Breaking, \\
Negative Metric and Vacuum Energy\footnote{%
Based on the talk presented at the 6th CST-MISC Joint Symposium on 
Particle Physics, Oct.~15-16, 2016, Campus Plaza Kyoto, Kyoto, Japan.  
}%
}%
\author{%
Taichiro Kugo
}{%
Department of Physics and
Maskawa Institute for Science and Culture, \\
Kyoto Sangyo University, Kyoto 603-8555, Japan
}%
\abstract{The supersymmetric Nambu-Jona-Lasinio model proposed by 
Cheng, Dai, Faisel and Kong is re-analyzed by using an auxiliary 
superfield method in which a hidden local U(1) symmetry emerges. 
It is shown that, 
in the healthy field-space region where no negative metric particles appear, 
only SUSY preserving vacua can be realized in the weak coupling regime and 
a composite massive spin-1 supermultiplets appear as a result of spontaneous 
breaking of the hidden local U(1) symmetry. In the strong coupling regime, 
on the other hand, SUSY is dynamically broken, but it is always 
accompanied by negative metric particles. }
\vspace{5mm}
\section{Introduction}

Global supersymmetry (SUSY) implies the non-negative definite vacuum energy. 
This indeed follows from the SUSY algebra
\begin{equation}
\{\,Q_\alpha, \ \bar Q_{\dot\beta}\,\} = 2(\sigma_\mu)_{\alpha\dot\beta}P^\mu 
\end{equation}
and the assumption of the positive metric of the state vector space;  
that is, using $\bar Q_{\dot\alpha}=(Q_\alpha)^\dagger$, we have 
\begin{equation}
\braket<0|H|0> = 
\frac14 \sum_{\alpha,\dot\alpha=1,2} 
\braket<0|(\bar Q_{\dot\alpha})^\dagger\bar Q_{\dot\alpha}+Q^\dagger_\alpha Q_\alpha|0>
=\frac14 \sum_{\alpha,\dot\alpha=1,2} \left(
\norm{\,\bar Q_{\dot\alpha}\ket{0}}^2 + \norm{\,Q_\alpha\ket{0}}^2\right) \geq0.
\end{equation}
The VEV of the Hamiltonian could be negative only when the 
zero-momentum {\em Goldstino} states  
$\,Q_\alpha\ket{0}, \ \bar Q_{\dot\alpha}\ket{0}$ have {\em negative metric}, 
which would, however, imply the disaster for the theory. 

Thus the potential energy of the vacuum is bounded from below and the 
minimum value zero is saturated by the normal SUSY vacuum. 
Therefore, the dynamical breaking of SUSY, or any other symmetries, 
is generally very {\em difficult} in the globally supersymmetric theory. 

Well-known examples realizing the spontaneous SUSY breaking are 
Fayet-Illiopoulos\cite{Fayet:1974jb} and Fayet-O'Raifeartaigh\cite{Fayet:1975ki}\cite{ORaifeartaigh:1975nky} models; the vanishing conditions of the 
D- and F-terms (in the former), or various F-terms (in the latter) 
in the tree-level potential cannot simultaneously 
be satisfied so that SUSY stationary points do not exist. 
More dynamical SUSY breaking models are also known but they are rather 
implicit and highbrow gauge theory models\cite{Izawa:1996pk}\cite{Intriligator:1996pu}\cite{Izawa:2009nz}, 
some of which are shown dual to the Fayet-O'Raifeartaigh models 
via Seiberg's duality\cite{Intriligator:1995au}.

It is, therefore, highly desirable to find an explicit simple model which 
exhibits dynamical SUSY breaking. If such a tractable model is found, 
it would have much utility in phenomenological models for generating 
the soft SUSY breaking terms dynamically\footnote{
Ohta and Fujii have shown that all the soft SUSY breaking terms can be 
generated by a kind of spontaneous SUSY breaking by 
`dipole ghost mechanism'\cite{Ohta:1981bi,Ohta:1982ys} They there discussed 
also the connection between the positivities of the vacuum energy and of the 
Goldstino's norm. }, 
just as the old Nambu-Jona-Lasinio 
model has long been used as a semi-quantitative parallel model for the 
dynamical chiral symmetry breaking realized in QCD\cite{Hatsuda:1994pi}. 

%

Recently,  
Cheng, Dai, Faisel and Kong (CDFK)\cite{Cheng:2015dgt,Cheng:2016dvq}
%
proposed a {\em supersymmetric NJL model} which, they claim, realizes 
the dynamical SUSY breaking. This model is indeed very simple one. So, if 
it is really a healthy model suffering no 
negative metric problem, it is very important and will become the desired 
useful model of dynamical SUSY breaking. 

Their paper is, however, not written in a crystalclear way. We therefore 
re-analyze their model in this paper, and make clear what actually happens 
there, in particular, from the viewpoint of the positive/negative metric 
problem of the particles. 


\section{Supersymmetric NJL model by Cheng-Dai-Faisel-Kong}


\subsection{The massless model and an equivalent auxiliary field model with hidden $U(1)$ symmetry}

The supersymmetric NJL-like model considered by Cheng-Dai-Faisel-Kong (CDFK) 
reads 
\begin{equation}
\shbox{$\displaystyle 
{\cal L}=\int d^4\theta \Bigl(\bar\Phi\Phi - {G\over2N}\bigl(\bar\Phi\Phi \bigr)^2 \Bigr), 
\qquad \bar\Phi\Phi \equiv\sum_{i=1}^N \bar\Phi_i\Phi^i
$}
\label{eq:superNJL}
\end{equation}
where $\Phi^i=[ A^i,\ \psi^i,\ F^i ]$ is matter chiral superfields carrying the 
flavor index $i$ of SU(N). 
We analyze this model in the leading order in $1/N$ expansion. 
Actually, CDFK considered the massive model 
possessing the mass term 
\begin{equation}
{\cal L}_{\rm mass} = \int d^2\theta\ m\Phi\Phi  + {\rm h.c.}
\label{eq:mass}
\end{equation}
in which case the flavor symmetry reduces to SO(N). We first concentrate 
in this section on the simpler massless case, and defer the discussion for 
the massive case to the next section. 

Keep in mind that this model has a dangerous kinetic term like
\begin{equation}
\bigl(1-{G\over N}A^\dagger A\bigr)
\left(-\partial_mA^\dagger\cdot\partial^mA-i\bar\psi\bar\sigma^m\partial_m\psi\right)
\end{equation} 
which becomes of negative metric in the field space region $A^\dagger A>N/G$. 
If the realized vacuum point is well inside of this boundary, i.e., 
$\VEV{A^\dagger A}<N/G$, then this negative metric poses no problem.

Similarly to the usual non-SUSY NJL case, this model (\ref{eq:superNJL}) 
can also be 
equivalently rewritten by adding a Gaussian term of an auxiliary vector 
superfield $U$:\cite{Cheng:2015dgt,Cheng:2016dvq}
\begin{eqnarray}
{\cal L}&=&
\int d^4\theta \Bigl(\bar\Phi\Phi - {G\over2N}\bigl(\bar\Phi\Phi \bigr)^2 
+{N\over2G}\bigl(   U + {G\over N}\bar\Phi\Phi \bigr)^2 \Bigr) \nn
&=&\int d^4\theta \Bigl(\bar\Phi\Phi  \bigl(1+U \bigr)  
+{N\over2G} U^2 \Bigr) 
\label{eq:6}
\end{eqnarray}
This auxiliary superfield $U$ stands for the superfield pair $\bar\Phi\Phi $ 
by the equation of motion:
\begin{equation}
U =-{G\over N}\bar\Phi\Phi  \qquad \hbox{so that \quad $\VEV{U}=0$ at $G=0$} 
\label{eq:7}
\end{equation}

Now, we rewrite this auxiliary vector superfield $U$, or the shifted one 
$U+1$ by 1, into 
\begin{equation}
U+1 = \bar\Sigma e^{2V} \Sigma  
\label{eq:8}
\end{equation}
by introducing a chiral superfield $\Sigma$ and a vector superfield $V$. 
This rewriting is, of course, {\em redundant} so that 
$U+1$ remains invariant under the following {\em hidden U(1)-gauge 
transformation} with a chiral superfield parameter $\Lambda$:
\begin{equation}
\left\{
\begin{array}{lcl}
\Sigma &\rightarrow& e^{-i\Lambda}\Sigma, \qquad \bar\Sigma\ \ \rightarrow\ e^{+i\bar\Lambda}\bar\Sigma,  \\
2V &\rightarrow& 2V+i(\Lambda-\bar\Lambda)   
\end{array}
\right.
\end{equation}
Of course, this gauge symmetry is {\em fake}, but it is very useful 
nevertheless. 
If we fix this gauge invariance by taking an `axial gauge' $\Sigma=1$, 
then this is merely 
an equivalent rewriting $U\ \rightarrow\ V=\ (1/2)\ln(U+1)$ of vector superfield 
variable. 
But we can take any other gauges which must be gauge-equivalent with one
another. We shall take Wess-Zumino gauge below. 

Further, if we redefine the original chiral matter $\Phi^i$ into $\phi^i$ as
\begin{equation}
\Sigma\Phi^i \equiv \phi^i, \qquad 
\bar\Sigma\bar\Phi^i \equiv \bar\phi^i, 
\label{eq:10}
\end{equation}
then, the Lagrangian (\ref{eq:6}) becomes
\begin{equation}
\shbox{$
\displaystyle 
{\cal L}=\int d^4\theta \Bigl(\bar\phi_ie^{2V}\phi^i 
+{N\over2G} \bigl( \bar\Sigma e^{2V} \Sigma -1\bigr)^2 \Bigr) 
$}
\end{equation}
which is hidden U(1)-gauge invariant under
\begin{equation}
\left\{
\begin{array}{lcl}
\phi^i &\rightarrow& e^{-i\Lambda}\phi^i, \qquad \bar\phi_i\ \ \rightarrow\ e^{+i\bar\Lambda}\bar\phi_i,  \\
\Sigma &\rightarrow& e^{-i\Lambda}\Sigma, \qquad \bar\Sigma\ \ \rightarrow\ e^{+i\bar\Lambda}\bar\Sigma,  \\
2V &\rightarrow& 2V+i(\Lambda-\bar\Lambda)   
\end{array}
\right.
\end{equation}
Using this, we can take the Wess-Zumino gauge in which
\begin{equation}
V = \left[\ C,\, Z_\alpha,\, H,\, K,\, v_m,\, \lambda,\, -D\ \right] \ \rightarrow\ 
\left[\ 0,\, 0,\, 0,\, 0,\, v_m,\,\lambda,\, -D\ \right] 
\end{equation} 
and $\Sigma$ becomes a normal chiral `matter':
\begin{equation}
\Sigma = \left[\ z,\ \chi,\ h\ \right] . 
\end{equation}
Note that we are taking {\em negative} sign convention for the 
$D$ field of the vector multiplet $V$ for later convenience. 

\subsection{Effective potential in the leading order in $1/N$}

We use the covariant derivative $\calD_m$ which is defined to be 
$\calD_m\equiv\partial_m +iqv_m$ on every component 
fields $\varphi_q$ of chiral superfields with $U(1)$-charge $q$, 
transforming $\varphi_q\rightarrow e^{-iq\Lambda}\varphi_q$. Then the 
part of the Lagrangian for 
the chiral matter field $\phi^i=[\,A^i,\ \psi^i,\ F^i\,]$ with $q=1$ reads
\begin{equation}
\int d^4\theta\ \bar\phi_ie^{2V}\phi^i
=
\bigl(A_i^\dagger \ \bar\psi_i \ F_i^\dagger\bigr)
\underbrace{
\begin{pmatrix}
\calD_m\calD^m-D & \sqrt2 i\lambda & 0 \\
-\sqrt2 i\bar\lambda & -i\bar\sigma^m\calD_m & 0 \\
0 & 0& 1
\end{pmatrix}
}_{\equiv\,\Delta}
\begin{pmatrix}
A^i \\ \psi^i \\ F^i
\end{pmatrix}
\label{eq:15}
\end{equation}
aside from the total derivative terms. 
In the leading order in $1/N$ expansion, the effective action $NS$ is given by
\begin{equation}
N S = N \left[
+i\,{\rm STr}\,{\rm Ln} (-\Delta)
+\int d^4x d^4\theta {1\over2G} \bigl( \bar\Sigma e^{2V} \Sigma -1\bigr)^2 
\right]
\label{eq:16}
\end{equation}
with ${\rm STr}$ denoting the functional supertrace. How this supertrace 
term can be evaluated diagrammatically is explained in the Appendix. 
Noting that only the bosonic scalar fields can take the constant 
($x$-independent) VEV's 
\begin{equation}
\VEV{\Sigma}= [\,z,\ 0,\ h\,], \qquad 
\VEV{V}= [\,0,\ 0,\ -D\,]
\end{equation}
and inserting these VEV's into the action (\ref{eq:16}), we find the 
effective potential $V$ given by 
\begin{equation}
V(z,h,D)= \int^\Lambda{d^4k\over(2\pi)^4} \left( \ln (k^2 +D) - \ln (k^2) \right)
-\int d^4\theta\ {1\over2G} \Bigl( \VEV{\bar\Sigma} e^{2\VEV{V}} \VEV{\Sigma} -1\Bigr)^2.
\end{equation}
(Note that the true potential is $N$ times this  $V$.) 
Here $k^\mu$ denotes the {\em Euclidean} 4-momentum. 
As is usual in NJL model, this one-loop integral is divergent so we put 
the ultra-violet momentum cut-off $\Lambda$ on 
the Euclidean 4-momentum integration as $k^2\leq\Lambda^2$.  
Performing the momentum integration $\int^\Lambda d^4k$ 
and Grassmann integration 
$d^4\theta$, we finally obtain the explicit form of effective potential in the 
$1/N$ leading order:
\begin{equation}
\shbox{$
\begin{array}{rcl}
G\,V(z,h,D)&=& \displaystyle {G\over32\pi^2}\Bigl[
\Lambda^4\ln(1+{D\over\Lambda^2})-D^2\ln(1+{\Lambda^2\over D})+D\Lambda^2\Bigr] \\[3ex]
&& {}+(1-2\abs{z}^2)\abs{h}^2+ (\abs{z}^2-1)\abs{z}^2D 
\end{array}
$}
\label{eq:pot}
\end{equation}

Stationarity conditions of this potential lead to
\begin{eqnarray}
{\delta V\over\delta h}=0 &\Rightarrow& (2\abs{z}^2-1) h^*=0 \ \Rightarrow\ h=0 \hbox{\ or\ } \abs{z}^2=1/2 
\label{eq:20}\\
{\delta V\over\delta z}=0 &\Rightarrow& \bigl[2\abs{h}^2-(2\abs{z}^2-1)D\bigr] z^*=0 \nn
&\Rightarrow&\ h=0 \hbox{\ and\ } (D=0 \hbox{\ or\ } \abs{z}^2=1/2) 
\label{eq:21}\\
{\delta V\over\delta D}=0 &\Rightarrow& 
{G\over32\pi^2}\Bigl[2\Lambda^2-2D\ln(1+{\Lambda^2\over D})\Big]
= (1-\abs{z}^2)\abs{z}^2 
\label{eq:22}
\end{eqnarray}
In (\ref{eq:21}), we have excluded the possibility $z^*=0$. This is because 
the first 
component of $U+1$, being proportional to $\abs{z}^2$ by 
Eq.~(\ref{eq:8}) and giving the coefficient of the kinetic term of the 
matter field $\Phi^i$ by Eq.~(\ref{eq:6}), should not vanish.
From Eqs.~(\ref{eq:20}) and (\ref{eq:21}), $h$ must vanish in any case and 
$D=0$ or $\abs{z}^2=1/2$. 

The right hand side (RHS) of Eq.~(\ref{eq:22}), $(1-\abs{z}^2)\abs{z}^2$, 
takes the value between $0\leq \hbox{RHS}\leq1/4$ for 
$1\geq\abs{z}\geq0$: At free theory limit $G=0$, Eqs.~(\ref{eq:7}) and (\ref{eq:8}) 
implies 
\begin{equation}
\abs{z}^2 =1, \ \ \ \hbox{so that, by Eq.~(\ref{eq:21}),} \ \ \ D=0.
\end{equation}
As $G$ becomes larger starting from 0, 
$\abs{z}^2$ becomes smaller from 1 until it reaches the point 
$\abs{z}^2=1/2$. Until then, the stationary point has to keep $D=0$ because of 
Eq.~(\ref{eq:21}) so that the value $\abs{z}^2$ is determined by 
Eq.~(\ref{eq:22}) as
\begin{equation}
{G\over32\pi^2}\,2\Lambda^2= (1-\abs{z}^2)\abs{z}^2 .
\end{equation} 
This continues until the coupling constant $G$ reaches the critical value 
$G^0_{\rm cr}$ where $\abs{z}$ comes down to the point 
$\abs{z}^2=1/2$ realizing the maximum $1/4$ of 
the RHS $(1-\abs{z}^2)\abs{z}^2$: 
\begin{equation}
{G^0_{\rm cr}\over32\pi^2}2\Lambda^2= {1/4} \ \ \Rightarrow\ \ 
G^0_{\rm cr}=4 \pi^2/\Lambda^2.
\end{equation}
When $G$ further becomes larger beyond this critical value, 
$\abs{z}^2$ stays at this maximum point of the RHS, $\abs{z}^2=1/2$, so that 
$D$ can no longer be zero, as determined by Eq.~(\ref{eq:22}):
\begin{equation}
{D\over\Lambda^2}\ln(1+{\Lambda^2\over D})= 1- {G^0_{\rm cr}\over G}
\label{eq:26}
\end{equation}
That is, the SUSY is {\em spontaneously broken}. 

Note that 
{\em there is no stationary point which keeps SUSY} in this strong 
coupling region $G>G^0_{\rm cr}$. 
Indeed, the SUSY points realizing $D=0$ and $h=0$ surely 
realizes vanishing value of the potential $V(z, h, D)$ in (\ref{eq:pot}) 
and the stationarity $\partial V/\partial z=0$ and 
$\partial V/\partial h=0$ with respect to $z$ 
and $h$, as Eqs.~(\ref{eq:21}) and C(\ref{eq:20}) show, 
independently of the $G$-value. However, any SUSY points on the line 
$D=0$ (with $h=0$), have {\em non-vanishing gradient} $\partial V/\partial D|_{D=0}\not=0$ 
for $G>G^0_{\rm cr}$ since
\begin{eqnarray}
{\partial V\over\partial D}\bigg|_{D=h=0}&=&
{G\over32\pi^2}\Bigl[2\Lambda^2-2D\ln(1+{\Lambda^2\over D})\Bigr]\bigg|_{D=0}
-(1-\abs{z}^2)\abs{z}^2 \nn
&=&{G\over32\pi^2}\,2\Lambda^2-(1-\abs{z}^2)\abs{z}^2 
\geq\frac14\biggl(
{G\over G^0_{\rm cr}}-1\biggr) >0\,,
\end{eqnarray}
because of $(1-\abs{z}^2)\abs{z}^2\leq1/4$, so that 
they cannot be the vacuum candidate. 

\subsection{What about negative metric?}

The above analysis shows that the SUSY is really 
spontaneously broken dynamically for the strong coupling 
region $G>G^0_{\rm cr}$ in this model. 
However, this model had a {\em potential danger of negative 
metric particles} which might appear depending on the vacua characterized 
by VEV's of the fields. 
So let us examine it. 

The original chiral supermultiplets $\Phi^i$ are redefined into $\phi^i$ 
which possesses ordinary gauge-invariant kinetic term $\bar\phi_ie^{2V}\phi^i$ 
and have no more complicated interaction in this leading order in $1/N$. 
So they remain to have {\em positive metric} irrespectively of the VEV's 
of the fields $D$, $h$ and $z$.

Therefore, we have only to analyze the kinetic term of the 
vector multiplet $V=[\lambda, v_m, D]$ and the chiral matter $\Sigma=[z, \chi, h]$. 
First, the vector multiplet is described by a slightly complicate-looking 
`kinetic term' $i\,{\rm STr\,Ln}(-\Delta)$ in the action (\ref{eq:16}) in the 
leading order in $1/N$, but it actually appears as a mere one-loop diagrams
of the `healthy' chiral multiplet fields $\phi^i$ possessing normal minimal 
gauge coupling $\bar\phi_ie^{2V}\phi^i$ as seen in (\ref{eq:15}), 
and thus, there is no reason for the negative metric to appear for the 
vector multiplet fields $V=[\lambda, v_m, D]$. 

Thus, we have only to consider the problem only for the chiral matter 
$\Sigma=[z, \chi, h]$. To see this we need an explicit component field expression 
for the term $\int d^4\theta(\bar\Sigma e^{2V} \Sigma -1)^2$ in the leading order 
action (\ref{eq:16}), which can be obtained most easily as done in the 
Appendix:
\begin{eqnarray}
&&\hspace{-2em}
\int d^4\theta\, \bigl( \bar\Sigma e^{2V} \Sigma -1\bigr)^2  \nn
&&{}=2(2\abs{z}^2-1)\Bigl( \abs{h}^2 - 
\calD_mz^*\calD^mz
-{i\over2}\bar\chi\bar\sigma^m\overleftrightarrow{\calD}_m\chi + 
\sqrt2 i (z^*\lambda\chi-z\bar\lambda\bar\chi)
\Bigr)
\nn
&&{}-2\abs{z}^2(\abs{z}^2-1)D -2i(\bar\chi\bar\sigma^m\chi)(z^*\overleftrightarrow{\calD}_mz)
-2(zh\bar\chi^2+z^*h^*\chi^2) +\chi^2\bar\chi^2
\label{eq:CompAction}
\end{eqnarray}
The first line give the kinetic terms of the fields $z$ and $\chi$. We note 
that those kinetic terms have a common field-dependent coefficient 
$(2\abs{z}^2-1)$. The value of this coefficient is not positive definite(!) 
and becomes {\em negative} if $\abs{z}^2$ becomes smaller than 1/2. 
So clearly, this model becomes disastrous in the region $\abs{z}^2<1/2$ 
owing to the appearance of negative metric modes.  

In the above, however, we have analyzed the effective potential $V(z,h,D)$ in 
Eq.~(\ref{eq:pot}) and found the stationary points at $1\geq\abs{z}^2>1/2$ 
with $h=0$ and $D=0$ for the weak coupling region $0\leq G< G^0_{\rm cr}$. 
So there are no problems in this weak coupling region.

However, as $G$ reaches the critical point $G^0_{\rm cr}$ or goes beyond, 
$\abs{z}^2$ exactly takes the value $1/2$! What   happens there? At first sight, 
it merely means that the kinetic terms of $z$ and $\chi$ disappear, or, they 
become of zero-norm particles. If these were true, then, there were no 
problems at all and the present CDFK model gives a simple healthy model 
exhibiting dynamical SUSY breaking. 

Unfortunately, however, a severe problem is there. The point is that the 
scalar field $z$ fluctuates around the VEV $\abs{\VEV{z}}=1/\sqrt2$ and the 
fluctuating scalar modes would have problematic indefinite metric. 
To see this, let us parametrize the scalar field $z$, taking the VEV be real, as
\begin{equation}
z(x) =  \frac1{\sqrt2}e^{i\theta(x)}\bigl( 1 +\phi(x) \bigr),
\end{equation}
then, $2\abs{z}^2-1= 2\phi+\phi^2$ and 
the scalar kinetic term in the above Lagrangian reads
\begin{equation}
2(2\abs{z}^2-1)\,\calD_mz^*\calD^mz
= (2\phi+\phi^2)\left(\partial_m\phi\,\partial^m\phi+v'_mv^{\prime m}(1+\phi)^2\right)
\end{equation}
Here the NG boson field $\theta(x)$ of the spontaneously broken U(1) was absorbed 
into the vector field $v'_m=v_m+\partial_m\theta$.
The kinetic term of $\phi$, which looks like (since $\phi\gg\phi^2$ around $\phi\sim0$)
\begin{equation}
\phi\partial_m\phi\,\partial^m\phi 
\label{eq:30}
\end{equation}
is problematic since it has the {\em fluctuating (metric) sign} 
around $\phi=0$. When $\phi$ goes into the negative side $\phi<0$, $\phi$ becomes 
of negative metric, which is a disaster to the theory. 

If the kinetic term had a non-negative definite coefficient 
function $f^2(\phi)\geq0$, then it could be rewritten into a normal kinetic term by 
field redefinition:
\begin{equation}
f^2(\phi)\partial_m\phi\,\partial^m\phi = 
\partial_m\Phi\,\partial^m\Phi, \quad \hbox{with}\quad  \Phi= \int d\phi\,f(\phi)
\label{eq:Normalization}
\end{equation}
Note that the vanishingness itself of $f(\phi)$ at $\phi=0$ poses no problem; 
for instance, take the simplest example, $f(\phi)=\phi$, then $\Phi=\phi^2$. Important 
is the non-negative definiteness of the sign.

If we applied blind-mindedly this formula (\ref{eq:Normalization}) to
the present case (\ref{eq:30}), then we would have obtained a normal form of 
kinetic term for the new 
scalar field $\Phi=(2/3)\phi^{3/2}$. But, this also implies that 
the relation with the original scalar field $z$ is singular and actually the 
very point $\phi=0$ is a branch point; if $\phi$ goes into the negative side $\phi<0$, 
then $\Phi\propto\phi^{3/2}$ becomes purely imaginary, again implying the appearance of 
the disastrous negative metric.

Although we have discussed only the scalar modes up to here, the same problems 
occur also for the fermion mode, whose kinetic term reads
\begin{equation}
2(2\abs{z}^2-1)\,{-i\over2}\bar\chi\bar\sigma^m\overleftrightarrow{\calD}_m\chi 
=-i(2\phi+\phi^2)\,\bar\chi\bar\sigma^m\overleftrightarrow{\calD}_m\chi 
\end{equation}
This kinetic term also has the same problematic {\em fluctuating (metric) sign} 
around $\phi=0$. 
So this model becomes disastrous for the strong enough coupling region 
$G\geq G^0_{\rm cr}$ for which $\abs{\VEV{z}}^2=1/2$ is realized.

\subsection{Spontaneous breaking of hidden $U(1)$ gauge symmetry}

So as a model of dynamical SUSY breaking in the strong coupling region, 
the present model is unfortunately not a healthy model. 
However, the model always realizes $\abs{z}\not=0$ for $G>0$, so that 
the hidden U(1) gauge symmetry is always spontaneously broken. 
If we take $\VEV{z}$ real, then, ${\rm Im}\,z$ is the NG boson 
absorbed in the vector $v_m$ in $V=[\,v_m,\ \lambda,\ D\,]$.

For the weak coupling region $0<G< G^0_{\rm cr}$, the SUSY is not broken 
so that $({\rm Re}\,z,\ \chi)\in\Sigma$ form 
a {\em massive vector multiplet} $0\oplus 1/2\oplus 1$ with $(v_m,\ \lambda)\in V$. 
(But actually this massive vector multiplet is unstable.) Their mass terms 
appear in the first line of the action (\ref{eq:CompAction}) as
\begin{eqnarray}
=2(2\abs{z}^2-1)\Bigl( \abs{h}^2 - \underbrace{\calD_mz^*\calD^mz}
_{ \hbox{\small kinetic term of Re$z$}\atop + \hbox{\small vector mass} }
-{i\over2}\bar\chi\bar\sigma^m\overleftrightarrow{\calD}_m\chi + 
\sqrt2 i (\underbrace{z^*\lambda\chi-z\bar\lambda\bar\chi}_{\hbox{\small Dirac mass}})
\Bigr) 
\end{eqnarray}
As far as $G<G^0_{\rm cr}$, the VEV of the scalar $z$ gives $\abs{z}^2>1/2$ 
\footnote{Actually, the stationarity condition (\ref{eq:22}), which determines 
the value of VEV $\abs{z}$, is {\em symmetric} under the reflection of 
$\abs{z}^2$ around the point $\abs{z}^2=1/2$; i.e., 
$\abs{z}^2 \longleftrightarrow  1-\abs{z}^2$. So, the solution $\abs{z}^2>1/2$ 
for $G<G^0_{\rm cr}$, is always accompanied by another solution 
$\abs{z'}^2=1-\abs{z}^2<1/2$ which equally satisfies the stationarity of 
(\ref{eq:22}) and realizes even the same value of the potential 
$V(z, h=0, D)$. Those reflection solutions $\abs{z'}$ are, however, in the 
region $\abs{z'}^2<1/2$, where the kinetic term of $\Sigma$ is of negative metric 
there, and must be discarded.} 
so that the appearing mass terms of this vector multiplet carry correct 
signs. The kinetic terms for the hidden gauge boson and gaugino multiplet 
are dynamically generated by the one-loop diagrams $i {\rm STr\,Ln}(-\Delta)$ 
(whose expansion is given in Appendix). 
This massive vector multiplet for $G<G^0_{\rm cr}$ case does not give a true 
stable particles 
since the `constituent' matter multiplets $\phi^i=[\, A^i,\ \psi^i,\ F^i\,]$ 
are {\em massless} into which 
the vector multiplet can energetically decay. 
So in order to make the vector multiplet truly stable, 
it is necessary 
to put the mass term for the original matter multiplets $\Phi^i$. 
 

\section{Supersymmetric NJL model with massive constituent}

Let us now analyze the more complicated case; the CDFK NJL model which 
possesses the mass term (\ref{eq:mass}), $\int d^2\theta\ m\Phi\Phi  + {\rm h.c.}$,
of the original constituent chiral superfield $\Phi^i$. 
In the massless case 
we have used the auxiliary chiral $\Sigma$ and vector superfields $V$ and the 
matter fields $\phi^i$. Recall, however, that  
$\phi^i$ is not the original matter fields $\Phi^i$ but 
$\phi^i\equiv \Sigma\Phi^i$ in (\ref{eq:10}). 
In the massless case the matter fields appeared only in 
the combination of $\phi^i$, but here in massive case, the original matter 
fields $\Phi^i$ appear which are now written in a slightly complicated 
expression $\Phi^i=\Sigma^{-1}\phi^i$, and the mass term reads in terms of 
the component fields of $\phi^i=[A^i, \psi^i, F^i]$ and $\Sigma=[z, \chi, h]$:
\begin{eqnarray}
{\cal L}_{\rm mass} &=& \int d^2\theta\,\half\, m\Phi^i\Phi^i + {\rm h.c.}
= \int d^2\theta\, \half\,m\Sigma^{-2}\phi^i\phi^i + {\rm h.c.} \nn
&=& \int d^2\theta\, m\left(
z^{-2}A_iF_i-z^{-3}A_iA_ih -\half z^{-2}\psi_i\psi_i +z^{-3}A_i\chi\psi_i-\threehalf
z^{-4}A_iA_i\chi\chi \right) + {\rm h.c.}\nn
\label{eq:massComponent}
\end{eqnarray}
We take the mass parameter $m$ real and positive. 
Elimination of the auxiliary fields $F_i$ by the e.o.m. $F^*_i+mz^{-2}A_i=0$, 
yields the following `mass term' in the potential:
\begin{equation}
V= +m^2\abs{z}^{-4}A_i^\dagger A_i 
\end{equation}
Now, in the presence of the background fields $z, h, D$, the mass terms 
for the constituent bosons $A_i$ and fermions $\psi_i$ are given by
\begin{equation}
-(D+m^2\abs{z}^{-4})A_i^\dagger A_i-m\bigl(z^{-3}h\,A_iA_i
-\half z^{-2}\psi_i\psi_i+{\rm h.c.}\bigr)
\end{equation}
The mass square eigenvalues are given by 
\begin{eqnarray}
&&A_i: \ \ D+M^2 \pm mH, \qquad 
\psi_i: \ \  M^2, \nn
&&\hbox{with}\quad 
H\equiv2\abs{h}\abs{z}^{-3}, \quad M^2\equiv m^2/\abs{z}^4
\end{eqnarray}
Therefore the effective potential (\ref{eq:pot}) in the leading order in 
$1/N$, is now replaced by the 
\begin{equation}
\shbox{$
\begin{array}{rcl}
32\pi^2 V(z,h,D)&=& F(D+M^2+mH) + F(D+M^2-mH) - 2F(M^2) \\[1.5ex]
&&{}+\displaystyle\frac{32\pi^2}{G}\left[-\quarter (2\abs{z}^2-1)H^2\abs{z}^6
+(\abs{z}^2-1)\abs{z}^2D \right]
\end{array}
$}
\label{eq:m-pot}
\end{equation}
in terms of the function 
\begin{equation}
F(x)\equiv16\pi^2\int^{\Lambda}{d^4k\over(2\pi)^4}[\ln(k^2+x)-\ln(k^2)]=
\frac12\left[
\ln(1+x)-x^2\ln(1+\frac1x)+x \right].
\end{equation}
In this expression and hereafter, we set the cutoff $\Lambda$ equal to 1 for 
simplicity of writing. It would easily be recovered, if necessary, by 
considering the dimension. 
Note that the potential (\ref{eq:m-pot}) clearly reduces when $m=0$ to 
the previous expression (\ref{eq:pot}) ($\times 32\pi^2/G$) for massless case.

The analysis of the stationary points of this potential for massive case 
becomes rather complicated and difficult to do systematically, but was 
performed by CDFK rather extensively. They found several peculiar 
stationary points realizing $h\not=0$, which are usually difficult to 
understand the behavior physically. 

As far as we have analyzed, those peculiar stationary points seem to realize 
{\em negative} vacuum energy, thus suggesting the existence of the negative 
norm states. Indeed the examples we checked corresponded to the vacua 
realizing the scalar VEVs $\abs{z}^2 < 1/2$. However, the kinetic term 
of the chiral multiplet $\Sigma=[z,\chi,h]$ is still given by 
(\ref{eq:CompAction}) possessing the coefficient $(2\abs{z}^2-1)$. 
Therefore, for the theory to give a healthy theory, 
the stationary points should exist in the region 
\begin{equation}
\abs{z}^2> 1/2.
\end{equation}
The stationary points in the region $\abs{z}^2\leq1/2$ correspond to the vacua 
on which negative norm chiral multiplet particles appear so that they 
must be discarded. 

So here we analyze the potential restricting only in the healthy 
region $\abs{z}^2> 1/2$, and shall prove that 
there appears no dynamical SUSY breaking vacua, unfortunately. This 
conversely implies that many dynamical SUSY breaking solutions found by 
CDFK exist only in the unhealthy region $\abs{z}^2\leq1/2$.  

As three independent variables of the effective potential $V$, let us 
take the variables $D, H=2\abs{h}\abs{z}^{-3}, Z\equiv\abs{z}^2$ 
in place of the original ones $D, h, z$.
Stationarity conditions of the potential (\ref{eq:m-pot}) are given by
\begin{eqnarray}
{\delta V\over\delta D}=0 &\Rightarrow& 
I(D+M^2+mH)+I(D+M^2-mH)
= \frac{32\pi^2}G Z(1-Z) 
\label{eq:41}\\
{\delta V\over\delta H}=0 &\Rightarrow& m\Bigl(I(D+M^2+mH)-I(D+M^2-mH)\Bigr)
= \frac{16\pi^2}G (2Z-1)Z^3H
\label{eq:42}\\
{\delta V\over\delta Z}=0 &\Rightarrow& 
\Bigl(I(D+M^2+mH)+I(D+M^2-mH)-2I(M^2)\Bigr)(-2m^2Z^{-3}) \nn
&&{}=
\frac{32\pi^2}G \bigl[{H^2\over4}Z^2(8Z-3)+(1-2Z)D\bigr] 
\label{eq:43}
\end{eqnarray}
where
\begin{equation}
I(x)=\frac{d}{dx}F(x)=16\pi^2\int^\Lambda{d^4k\over(2\pi)^4}\frac1{k^2+x}=
1 - x\ln(1+\frac1{x}) 
\end{equation}
First of all, from the stationarity condition (\ref{eq:42}) with respect to 
$H (=2\abs{h}\abs{z}^{-3})$ which must be non-negative by definition, 
we immediately see that 
$H$ must vanish. $H=0$ clearly satisfies (\ref{eq:42}), but any positive 
$H$ cannot satisfy it. This is because 
$I(x)$ is a monotonically decreasing function from 1 to 0 for 
$0\leq x<\infty$. Therefore, if $H>0$, 
\begin{equation}
I(D+M^2+mH)-I(D+M^2-mH)<0 \quad \hbox{while} \quad (2Z-1)Z^3H>0,
\end{equation}
so that both sides of (\ref{eq:42}) have opposite sign in the 
present healthy region $1/2<Z<1$. 

This fact $H=0$ is the reason why the potential analysis remains very 
simple in this healthy region. 
Then, with $H=0$, the stationarity (\ref{eq:41}) with respect to $D$ 
gives the condition
\begin{equation}
I(D+m^2Z^{-2})=\frac{16\pi^2}G Z(1-Z).
\end{equation}
SInce $I(x)$ is monotonically decreasing from 1 to 0 as $x$ 
goes from 0 to $\infty$, the solution $D=D(Z)$ to this equation is formally 
written in the form
\begin{equation}
D(Z)= I^{-1}\Bigl(\frac{16\pi^2}G Z(1-Z)\Bigr)-m^2Z^{-2} \quad \hbox{for}\quad 
1>\frac{16\pi^2}G Z(1-Z)\geq0
\end{equation}
$I^{-1}(x)$ is the inverse function of $I$ existing for $0<x<1$; 
$I^{-1}(0)=+\infty$ and $I^{-1}(1)=+0$. $I^{-1}(x)$ is also monotonically 
decreasing function and $D(Z)$ is monotonically increasing function of $Z$. 
When $Z$ approaches the free point $Z=1$, $D(Z)$ becomes very large and 
behaves like 
\begin{equation}
D(Z) \sim \frac{G}{32\pi^2}\frac1{1-Z}\quad \hbox{as}\quad Z \rightarrow 1.
\end{equation}
If $G$ is smaller than the previous critical value $G^0_{\rm cr}$ for the 
massless case,
\begin{equation}
G<G^0_{\rm cr}\equiv4\pi^2\quad \rightarrow\quad     \frac{16\pi^2}G \frac14> 1,
\end{equation}
as $Z$ comes down from 1 towards 1/2, the argument $(16\pi^2/G) Z(1-Z)$ reaches 
the value 1 at some $\exists Z_0>1/2$ before $Z=1/2$; $(16\pi^2/G) Z_0(1-Z_0)=1$. 
That is, the stationary point $D(Z)$ of the potential 
with respect to $D$
exists only for the region $Z_0\leq Z\leq1$, but no stationary point 
for $1/2\leq Z<Z_0$. For $G>G^0_{\rm cr}$, the stationary points $D(Z)$ always 
exist for the whole healthy region $1/2\leq Z\leq1$. 

Note that even the coupling constant $G$ reaches the massless critical 
value $G^0_{\rm cr}$, the stationary point $D(Z)$ at $Z=1/2$ is negative 
$D(Z=1/2)=-m^2/4<0$ so that the $D(Z)=0$ is realized at some $Z>1/2$. 
A massive case critical coupling $G^m_{\rm cr}$ may be defined to be 
the coupling constant at which $D(Z=1/2)=0$ is realized; that is,
\begin{equation}
I(m^2/4) = 4\pi^2/G^m_{\rm cr},\qquad \hbox{so}\quad G^m_{\rm cr}>G^0_{\rm cr}. 
\end{equation}
Below this critical point $G<G^m_{\rm cr}$, the $D(Z)=0$ stationary 
point always exist in the healthy region $1/2<Z<1$. 

Finally consider the stationarity condition (\ref{eq:43})
with respect to $Z$, which reads now setting $H=0$
\begin{equation}
I(D+M^2)-I(M^2) 
=
\frac{8\pi^2}G \frac{Z^3}{m^2}(2Z-1)D
\label{eq:52}
\end{equation}
If we put $D=D(Z)$, then, this equation determines the stationary point $Z$.
This equation is very similar to the previous $H$-stationarity condition 
(\ref{eq:42}). If $D$ is positive, then the LHS of (\ref{eq:52}) is negative 
since $I$ is monotone decreasing, while the RHS is positive in the 
healthy region $1/2<Z<1$. If $D$ is negative, then the LHS is positive while 
the RHS is negative. So the only possibility for this equation to hold 
is $D=0$ case. We have seen above that 
the $D(Z)=0$ solution exists in the healthy region iff $G$ is 
below the critical 
coupling $G<G^m_{\rm cr}$. So in this weak coupling regime, 
there is a unique stationary point of the potential in the healthy region 
$1/2<Z<1$, and the SUSY remains unbroken since $H=D=0$. 

In the strong coupling region above the critical coupling 
$G>G^m_{\rm cr}$, on the other hand, 
$D$-stationary points $D(Z)$ are always positive $D(Z)>0$ in the healthy 
region, so that 
the $Z$-stationarity condition (\ref{eq:43}) cannot be satisfied. 
This means 
that there are no stationary points of the potential in the 
healthy region $1/2<Z<1$. 

Indeed, outside the healthy region, i.e., $Z<1/2$, 
one can find many stationary points which possess non-vanishing 
values of $H$ and/or $D$ and realize the negative vacuum energy values. 
Analysis is not easy. We, here, do not enter this problem, since 
whatever solutions may be found they necessarily represent 
unphysical theories.

\section{Conclusion}

We have re-analyzed an interesting supersymmetric NJL model proposed by 
Cheng-Dai-Faisel-Kong, by using a new type of auxiliary field method in which 
a hidden U(1) gauge symmetry emerges. Unfortunately, the model is shown to 
have no dynamical SUSY breaking vacua in the healthy field-space region 
in which no negative metric particles appear; if the coupling constant 
$G$ is weaker than a critical value $G^m_{\rm cr}$, the stationary point 
of the effective potential is uniquely SUSY vacuum, and if the coupling 
constant is larger than the critical value, $G>G^m_{\rm cr}$, there is no 
stationary point at all in the healthy field-space region. Above critical, 
the stationary points exist only in the un-healthy region so that the 
theory necessarily contain the negative metric particles. 

The model, therefore, represents healthy theory only in the weak coupling 
region $G<G^m_{\rm cr}$, where SUSY remains unbroken while the hidden 
U(1)-gauge symmetry is spontaneously broken and massive spin-1 
supermultiplet dynamically appear as composite particles.

\medskip 

The author would like to thank Otto Kong and Yifan Cheng for informing 
him of their work and valuable discussions. The discussions with Naoki Yamatsu and
Nobuyoshi Ohta are also acknowledged.

\newpage
\section{Appendix}
\subsection{Calculation of $\int d^4\theta(\bar\Sigma e^{2V}\Sigma-1)^2$}

From the Wess-Bagger\cite{Wess:1992cp}'s result for the chiral multiplet 
$\Sigma=[z, \chi, h]$ and 
the vector multiplet in the Wess-Zumino gauge, $V=[\lambda, v_m, -D]$, we have 
\begin{eqnarray}
\bar\Sigma e^{2V}\Sigma\Big|_{\theta^2\bar\theta^2} &=& 
\abs{h}^2-\abs{\calD_mz}^2 -\frac{i}2\bar\chi\bar\sigma^m\overleftrightarrow{\calD}_m\chi 
 +\sqrt2 i (z^*\lambda\chi-z\bar\lambda\bar\chi) -\abs{z}^2D\,.
 \label{eq:A1}
\end{eqnarray}
Then, the action for the square term $(\bar\Sigma e^{2V}\Sigma)^2$ is most easily 
obtained by applying this formula to the 
chiral superfield $\Sigma^2=[z^2, 2z\chi, 2zh-\chi^2]$ which carries the $q=2$ U(1) 
charge:
\begin{eqnarray}
\bigl(\bar\Sigma e^{2V}\Sigma\bigr)^2\Big|_{\theta^2\bar\theta^2} &=& 
(\bar\Sigma)^2e^{4V}(\Sigma)^2\Big|_{\theta^2\bar\theta^2}  \nn
&=& 
(2z^*h^*-\bar\chi^2)(2zh-\chi^2)-\calD_m(z^{*2})\calD^m(z^2)
-\frac{i}2(2z^*\bar\chi)\bar\sigma^m\overleftrightarrow{\calD}_m(2z\chi) \nn
&&
 +\sqrt2 i \left(z^{*2}(2\lambda)(2z\chi)-z^2(2\bar\lambda)(2z^*\bar\chi)\right) -\abs{z}^4(2D)
\label{eq:A2}
\end{eqnarray}
From Eqs.~(\ref{eq:A1}) and (\ref{eq:A2}), we find the expression cited in 
(\ref{eq:CompAction}): 
\begin{eqnarray}
\left(\bar\Sigma e^{2V}\Sigma-1\right)^2\Big|_{\theta^2\bar\theta^2}
&=&
2(2\abs{z}^2-1)\left(
\abs{h}^2-\calD_mz^*\cdot\calD^mz -\frac{i}2\bar\chi\bar\sigma^m
\overleftrightarrow{\calD}_m\chi+\sqrt2i(z^*\lambda\chi-z\bar\lambda\bar\chi)
\right) \nn
&&{}-2\abs{z}^2\left(\abs{z}^2-1\right)D
-2i(\bar\chi\bar\sigma^m\chi)(z^*\overleftrightarrow{\calD}_mz)
-2(zh\bar\chi^2+z^*h^*\chi^2)+\chi^2\bar\chi^2. \nn
\end{eqnarray} 

\subsection{
Expansion of the supertrace term $+i\,{\rm STr}\,{\rm Ln} (-\Delta)$}

\begin{eqnarray}
i\,{\rm STr}\,{\rm Ln} (-\Delta)
&=& 
i\,{\rm Tr}\,{\rm Ln} (-\calD_m\calD^m+D) 
-i\,{\rm Tr}\,{\rm Ln} 
\left(i\bar\sigma^m\calD_m-2\bar\lambda(-\calD_m\calD^m+D)^{-1}\lambda\right) \nn
&=& 
i\,{\rm Tr}\,{\rm Ln} (-\calD^2+D) 
-i\,{\rm Tr}\,{\rm Ln}(i\overline{\slash{\cal D}})
-i\,{\rm Tr}\,{\rm Ln}
\left(1 -2 \frac1{i\overline{\slash{\cal D}}}\,\bar\lambda\frac1{-\calD^2+D}\lambda\right) \nn
\end{eqnarray}
where in the second line we have used the notation
\begin{equation}
\calD_m\calD^m\equiv\calD^2,\qquad 
\bar\sigma^m\calD_m\equiv\overline{\slash{\cal D}}.
\end{equation}
The last term can be expanded as
\begin{equation}
-i\,{\rm Tr}\,{\rm Ln}
\left(1 -2 \frac1{i\overline{\slash{\cal D}}}\,\bar\lambda 
\frac1{-\calD^2+D}\lambda\right) \nn
=+i\sum_{n=1}^\infty\frac1{n}{\rm Tr}\,\left[
2(\lambda\,\frac1{i\overline{\slash{\cal D}}}\,\bar\lambda)
\frac1{-\calD^2+D}\right]^n
\end{equation}
Furthermore, when evaluating these action diagrammatically, we separate the 
free and interacting parts for $-\calD^2+D$ and $i\overline{\slash{\cal D}}$:
\begin{eqnarray}
-\calD^2+D&=& \underbrace{(-i\partial_m)^2+\VEV{D}}_{\Delta_A}+ 
\underbrace{(-i\partial_m)v^m + v^m(-i\partial_m) +v^2 +\tilde D}_{-{\cal L}_{\rm int}}
=\Delta_A-{\cal L}_{\rm int} \nn
i\overline{\slash{\cal D}}&=& 
i\overline{\slash{\partial}}-\overline{\slash{v}}
\end{eqnarray} 
with $\tilde{D} = D- \VEV{D}$. Then 
\begin{eqnarray}
i\,{\rm Tr}\,{\rm Ln} (-\calD^2+D)&=&
i\,{\rm Tr}\,{\rm Ln} (\Delta_A) 
-i\sum_{n=1}^\infty\frac1{n}{\rm Tr}\,\left[
\frac1{\Delta_A}\,{\cal L}_{\rm int}\right]^n \nn
-i\,{\rm Tr}\,{\rm Ln}(i\overline{\slash{\cal D}})
&=&
-i\,{\rm Tr}\,{\rm Ln} (i\overline{\slash{\partial}}) 
+i\sum_{n=1}^\infty\frac1{n}{\rm Tr}\,\left[
\frac1{i\overline{\slash{\partial}}}\,\overline{\slash{v}}\right]^n \nn
\frac1{-\calD^2+D}&=&\sum_{k=0}^\infty 
\frac1{\Delta_A}\left({\cal L}_{\rm int}\,\frac1{\Delta_A}\right)^k 
\nn
\frac1{i\overline{\slash{\calD}}}&=&\sum_{k=0}^\infty 
\frac1{i\overline{\slash{\partial}}}\left(\overline{\slash{v}}\  
\frac1{i\overline{\slash{\partial}}}\right)^k
\end{eqnarray}


\end{document}